\documentclass[10pt,letterpaper]{article}
\usepackage[top=0.85in,left=2.75in,footskip=0.75in,marginparwidth=2in]{geometry}

\usepackage[utf8]{inputenc}

\usepackage{cite}

\usepackage{nameref,hyperref}

\usepackage[right]{lineno}

\usepackage{microtype}
\DisableLigatures[f]{encoding = *, family = * }

\raggedright
\usepackage{changepage}

\usepackage[aboveskip=1pt,labelfont=bf,labelsep=period,singlelinecheck=off]{caption}

\makeatletter
\renewcommand{\@biblabel}[1]{\quad#1.}
\makeatother

\usepackage{lastpage,fancyhdr,graphicx}
\usepackage{epstopdf}
\fancyheadoffset[L]{2.25in}
\fancyfootoffset[L]{2.25in}

\usepackage{color}

\definecolor{Gray}{gray}{.25}

\usepackage{graphicx}

\usepackage{sidecap}

\usepackage{wrapfig}
\usepackage[pscoord]{eso-pic}
\usepackage[fulladjust]{marginnote}
\reversemarginpar

\newcommand{\footurl}[1]{\footnote{\url{#1}}}

\begin{document}
\vspace*{0.35in}

\begin{flushleft}
{\Large
\textbf\newline{The Transparent Relations Ontology (TRO): a vocabulary to describe conflicts of interest}
}
\newline
\\
Mikel Ega\~{n}a Aranguren \textsuperscript{1,*}
\\
\bigskip
\bf{1} Department of Computer Languages and Systems, University of Basque Country (UPV/EHU), Spain
\\
\bigskip
* mikel.egana@ehu.eus

\end{flushleft}

\section*{Abstract}
The Transparent Relations Ontology (TRO) offers a vocabulary to publish data about relations between powerful parties that should be more transparent, in order to detect possible conflicts of interest. TRO is based on minimal modelling, reusing common vocabularies to offer a simple yet useful resource to publish interoperable data about pointers to relations that might result in corruption cases. Additionally, best practices have been followed in order to sustain a technically rigorous ontology development process. A usage example with real data is mentioned, integrating information from Basque Government's Open Data services and a news outlet. Building upon its foundational design, future enhancements of TRO could significantly amplify its utility in uncovering and scrutinizing opaque relationships that may lead to corruption.


\section{Introduction}\label{intro}
Current democracies suffer a legitimacy crisis, since ruling bodies are regarded as ``extracting elites'' whose prime motivation is to enrich themselves rather than serve the public good \cite{68741}. Additionally, the cases in which this concern is grounded and corruption by political and economical elites does occur, the economic cost is considerable \cite{RePEc:imf:imfwpa:2019/253}. Therefore it is important that citizens and journalists are able to detect and document possible conflicts of interest that originate from the relation between ruling politicians and interested parties (Donation makers, companies, political lobbies, social organizations, etc.), in order to have healthier democracies. 

The data necessary to analyse those putative conflicts of interest is diverse, comprising, at least, contracts from public tenders, affiliation to political parties, personal relationships, corporation ownership, etc. The data is also sparse, isolated in silos, implemented in different formats, and often structured, semi-structured or unstructured, so the transparency activists need to build tailored and costly platforms to gather, integrate, and analyse the data of the domain in order to obtain interesting conclusions. An efficient solution to alleviate such technical problems is to have a common vocabulary to annotate the entities of the domain \cite{onto-integration}. Therefore, we present the Transparent Relations Ontology (TRO), an OWL\footurl{https://www.w3.org/TR/owl2-syntax/} vocabulary to represent conflicts of interest, in order to make the publication of data about corruption more FAIR (Findable, Accessible, Interoperable, Reusable \cite{wilkinson2016fair})  and ultimately help in its integration, processing and analysis.

The rest of the paper is organized as follows: Section \ref{sec:related} analyses related projects; Section \ref{sec:design} describes the modelling behind the ontology and a use case related to a Knowledge Graph (KG) built with currently available data; Section \ref{sec:technical} covers the technical details, including ontology development and publication; finally, Section \ref{sec:conclusion-future-work} provides wrapping conclusions and pointers for future work.

\section{Related work}\label{sec:related}
Currently there are no publicly available vocabularies that can be used to annotate data about conflicts of interest and corruption. There are projects that have published data about possible conflicts of interest through KGs in specific domains, like the Offshore Leaks Database\footurl{https://offshoreleaks.icij.org/} for the ``Panama papers'' or The Donation\footurl{https://ladonacion.es/} for the donations received by the former Spanish king Juan Carlos I. These are useful endeavours but both projects lack an explicit and public ontology: rather, they use implicit vocabularies that should be inferred by exploring the data, and cannot be reused to annotate other datasets, since they lack HTTP(S) resolvable and persistent URIs. This hinders their interoperability. 

Other projects tend to focus on only one side of the problem, namely, public tender processes: TheyBuyForYou \cite{DBLP:journals/semweb/SoyluCEBBMKPMTS22}, The Public Procurement Data Space (PPDS)\footurl{https://single-market-economy.ec.europa.eu/single-market/public-procurement/digital-procurement/public-procurement-data-space-ppds_en}, Tenders Guru\footurl{https://tenders.guru/}, Contratos Menores\footurl{https://lab.montera34.com/contratosmenores/}, Kontrata\footurl{https://github.com/erral/kontrata}, and more. Among those projects, TheyBuyForYou and PPDS stand out since they use the PPROC and eProcurement ontologies respectively: these ontologies are partly reused in TRO to represent contracts, but TRO offers additional elements to model, for example, the political affiliations or relations of a company owner, going beyond the tender process itself to model the interests behind it (In case they conform a conflict of interest). 

\section{Ontology design and usage}\label{sec:design}

\subsection{Modelling and reused ontologies}

The ontology design has been informed directly by the uses cases for publishing data, aiming at creating an immediately useful schema, rather than a complex axiomatic model that represents the whole knowledge domain. 

\begin{figure}[h]
    \vspace{2.5cm}
    \includegraphics[width=1\linewidth]{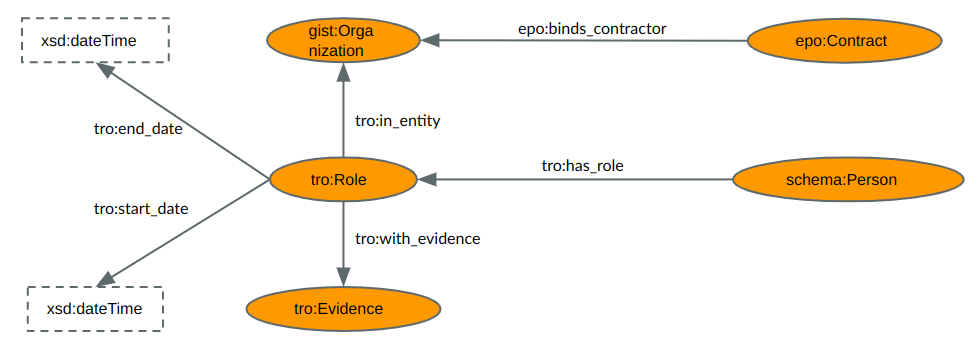}
    \caption{Basic modelling behind the TRO ontology: a person has a role during a certain time in a given organization, and such fact is backed by an evidence. The orange ovoids represent OWL classes; the arrows represent OWL restrictions; dashed boxes represent XSD Data Types. Not all the classes nor all the properties are shown: for all the entities of the ontology refer to Figure \ref{fig:entities}.}\label{fig:basic_model}
\end{figure}

\begin{figure}[h]
    \vspace{2.5cm}
    \includegraphics[width=1\linewidth]{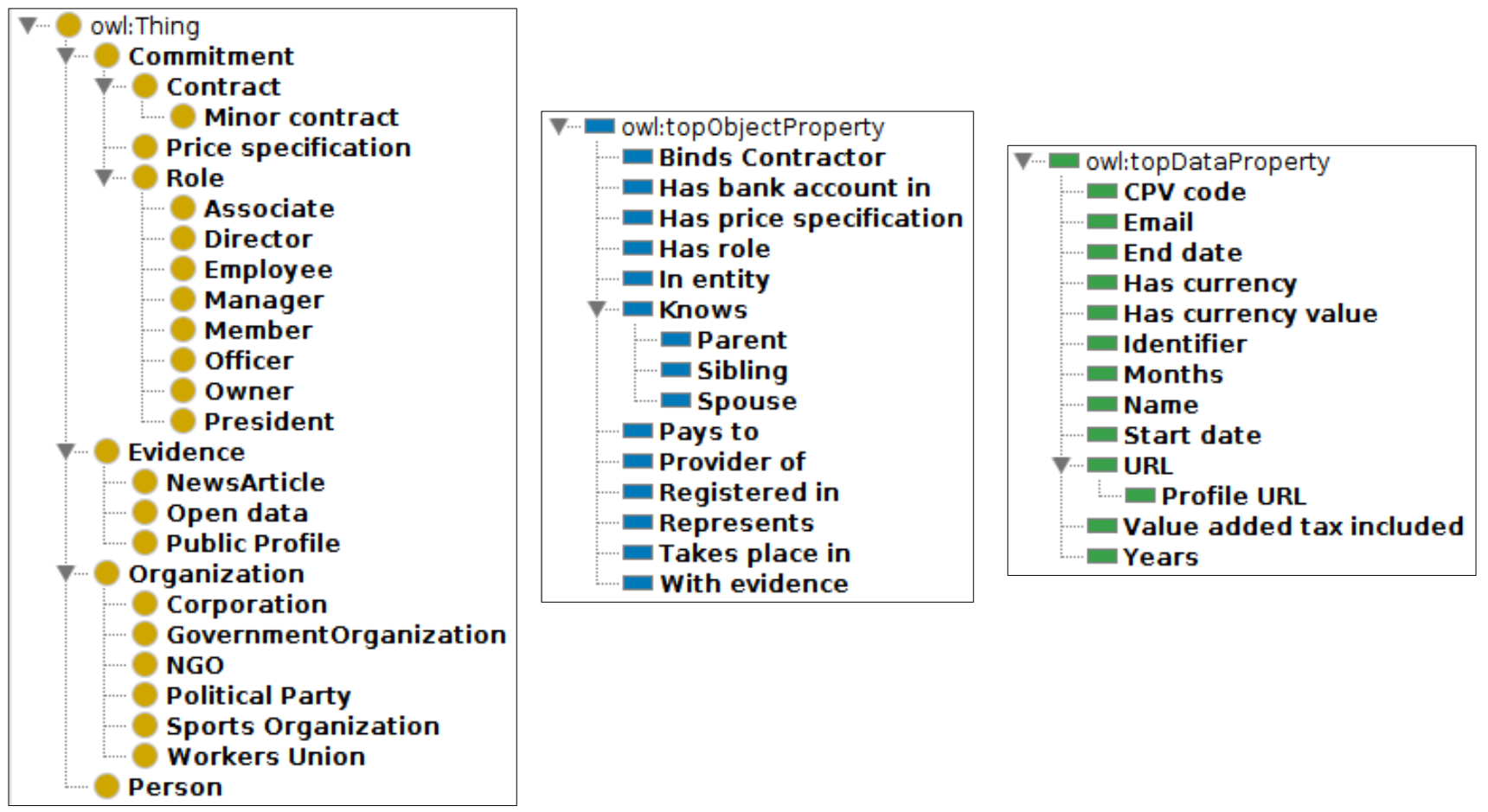}
    \caption{Entities of TRO, as shown in the Prot\'eg\'e interface. Hierarchies represent class/subclass relations and property/subproperty relations. Left: classes; center: object properties; right: data properties.}\label{fig:entities}
\end{figure}

The core of TRO is the idea that a person has a role in an entity, during a given time range, and such fact must be backed by an evidence (See Figure \ref{fig:basic_model}). That person might hold further personal or professional relations, affiliations to organisations, etc. that, when co-occurring with her role, might point to a possible conflict of interest. The main classes of such model and their descriptions are detailed in Table \ref{table:classes}. The rest of the entities of the ontology are summarised in Figure \ref{fig:entities}.  

TRO encompasses a fundamental upper ontology that serves as a foundation for interoperability, with the following disjoint classes: \texttt{Commitment}, \texttt{Organization}, \texttt{Evidence}, and \texttt{Person}. Also, entities from several external ontologies have been reused: GIST\footurl{https://github.com/semanticarts/gist}, Good Relations (GR) \cite{10.1007/978-3-540-87696-0_29}, Public Procurement Ontology (PPROC) \cite{10.3233/SW-150195}, eProcurement Ontology (ePO)\footurl{https://joinup.ec.europa.eu/collection/eprocurement/solution/eprocurement-ontology}, Schema\footurl{https://schema.org}, Time ontology\footurl{https://www.w3.org/2006/time} and DBpedia Ontology\footurl{https://dbpedia.org/ontology/}.

\begin{table}[]
    \begin{tabular}{lll}
        \hline 
     Class & Description & Origin  \\ \hline
     \texttt{Contract} & ``A voluntary, deliberate, and legally binding & epo \\ 
              & agreement between two or more competent parties'' &   \\ \hline
    \texttt{Organization}   & An organization (Corporation, Government Service,  & gist  \\ 
              & Union, etc.) \\ \hline
    \texttt{Evidence} & An evidence is a document that backs an statement  & tro \\
                      & (Usually the role of a person in an entity,  &     \\
                      & or the relation between people) and it must have       &     \\
                      & a URL. Evidences include: News Articles, Open Data      &    \\
                      & portals, public profiles, etc. This is not legal evidence & \\ \hline 
    \texttt{Person}   & A physical person with a compulsory name. She      & schema \\
                      & can have an email, an internet profile (e.g.    &  \\
                      & LinkedIn) etc. Every person has a role.         &  \\ \hline
    \texttt{Role}     & The function performed by a person in an entity,  \\ 
                      & during a given time, with an evidence & tro \\ \hline
    \end{tabular}
    \caption{Main classes of the TRO ontology. Left column: class names; center column: description; right column: ontology of origin (Prefix). Prefixes: tro: http://ehu.eus/tro\#, epo: http://data.europa.eu/a4g/ontology\#, gist: https://ontologies.semanticarts.com/gist/, schema: http://schema.org/.}\label{table:classes}
\end{table}

\subsection{Using TRO}\label{sec:usingTRO}
\begin{figure}
\vspace{2.5cm}
\includegraphics[width=1\linewidth]{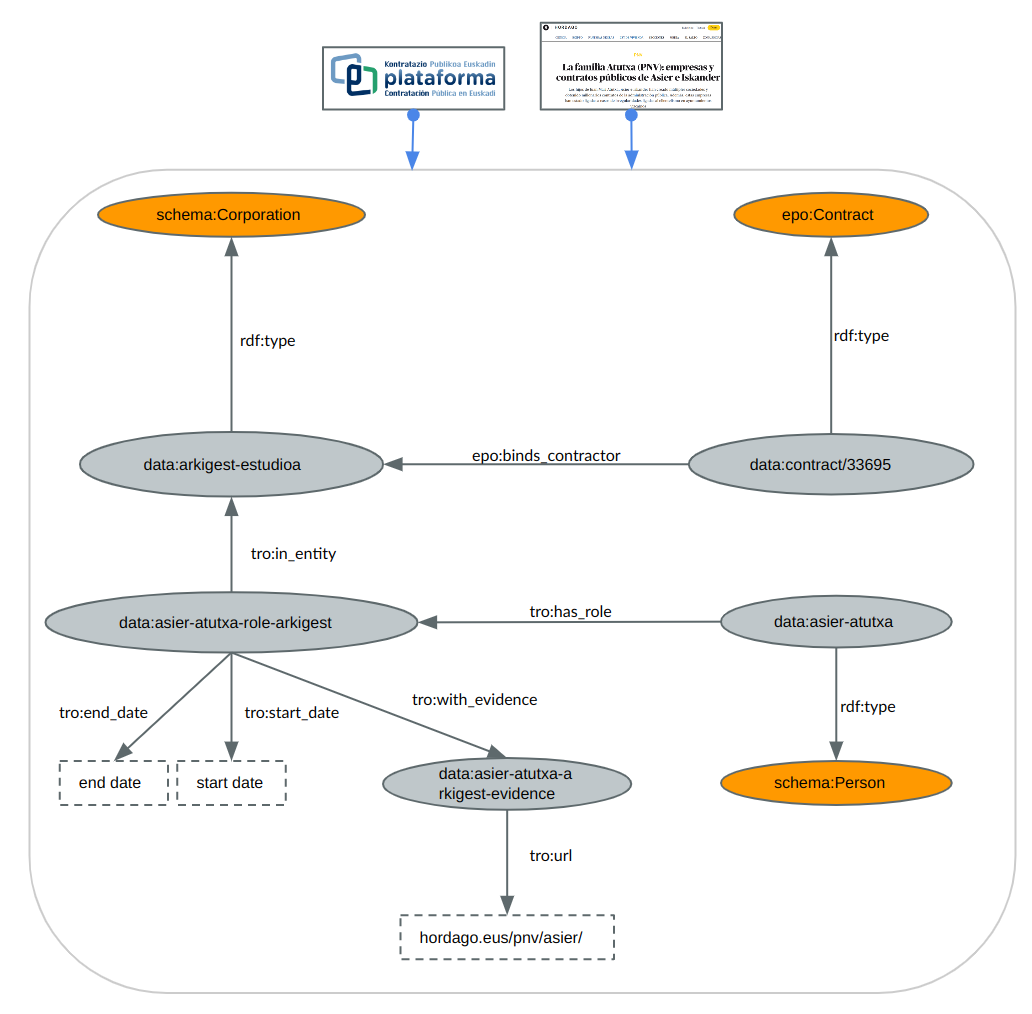}
\caption{Basque Country Institutions Transparent Relations Graph (BCITRG). The data is obtained from two sources (top): the Basque Government's tender registry, and a news source. The data is transformed to an RDF KG (Grey line encircling ovoids and arrows). A simplified sample is shown, and URIs are reduced for readability. Orange ovoids represent TRO classes; grey ovoids represent RDF nodes; dashed boxes represent literal values; TRO properties are used in all the RDF triples. The role of the person is identified by a unique URI composed by the person's normalised name, role type, dates of the role, and the organization, obtaining a completely unique URI for each role, on each date, on each organization, backed by an specific evidence. }\label{fig:bcitr}
\end{figure}

A use case for TRO can be found at the Basque Country Institutions Transparent Relations Graph (BCITRG) project\footurl{https://github.com/mikel-egana-aranguren/BasqueCountryInstitutionsTransparentRelationsGraph}. BCITRG's aim is ``to build a graph to integrate information about entities and individuals that might have a conflict of interest, in order to analyse such information''\footurl{https://github.com/mikel-egana-aranguren/BasqueCountryInstitutionsTransparentRelationsGraph/blob/main/README.md}. BCITRG integrates information from different sources into an RDF\footurl{https://www.w3.org/TR/rdf12-concepts/} KG. In this specific instance, information is obtained from the Basque Registry of Public Tenders\footurl{https://www.contratacion.euskadi.eus/} and Hordago\footurl{https://www.elsaltodiario.com/hordago/}, an investigative journalism outlet (Figure \ref{fig:bcitr}).  

\section{Technical information}\label{sec:technical}

\subsection{Development}

TRO is maintained in a GitHub repository\footurl{https://github.com/mikel-egana-aranguren/Transparent-Relations-Ontology} as a Turtle file\footurl{https://www.w3.org/TR/turtle/} generated by Prot\'eg\'e \cite{10.1145/2757001.2757003}. Its documentation is produced by the Widoco tool \cite{garijo2017widoco}.

The GitFlow branching model\footurl{https://nvie.com/posts/a-successful-git-branching-model/} and semantic versioning\footurl{https://semver.org/spec/v2.0.0.html} are followed as general methods, in order to maintain a sustainable and transparent development process. The \texttt{owl:priorVersion}, \texttt{owl:versionInfo} and \texttt{schema:schemaVersion} properties are used to describe the versioning. 

The ontology is validated in a Continuous Integration (CI) process\footurl{https://github.com/mikel-egana-aranguren/Transparent-Relations-Ontology/actions} that is triggered after every push to the develop branch. The CI process makes use of the ROBOT ontology engineering tool \cite{robot-owl}: ROBOT executes SPARQL queries to generate a quality report\footurl{http://robot.obolibrary.org/report} with differing log levels (ERROR, WARN, INFO) and also performs consistency checking through reasoning\footurl{http://robot.obolibrary.org/reason}. The SPARQL queries ROBOT executes check for quality in different dimensions, including compliance with the Linked Open Vocabularies metadata practices\footurl{https://lov.linkeddata.es/Recommendations_Vocabulary_Design.pdf} \cite{10.3233/SW-160213}. The CI process also generates an OQuaRE report \cite{duque2011oquare} that can be explored in the GitHub repository\footurl{https://github.com/mikel-egana-aranguren/Transparent-Relations-Ontology/tree/develop/oquare}. Finally, the ontology is regularly analysed in the OOPS service \cite{poveda2014oops}, and no pitfalls have been detected.

Provenance information is recorded for the ontology itself and all the terms that are added through the \texttt{dc:contributor}, \texttt{dcterms:created}, \texttt{dcterms:modified}, and \texttt{dc:date} properties.

\subsection{Availability}

The publication of TRO follows FAIR principles through the application of the methods suggested in \cite{https://doi.org/10.48550/arxiv.2003.13084}. A persistent URI for the ontology is served by the W3ID project\footurl{https://w3id.org/TRO}, including content negotiation, so that diverse agents can refer to it computationally. The W3ID URI redirects to the ontology files at GitHub\footurl{https://github.com/mikel-egana-aranguren/Transparent-Relations-Ontology}: the ontology itself is available as different OWL serialisations, altogether with HTML documentation. TRO can also be found at the LOV service\footurl{https://lov.linkeddata.es/dataset/lov/vocabs/tro} \cite{10.3233/SW-160213}. TRO is available under an Apache 2 License\footurl{https://apache.org/licenses/LICENSE-2.0} and its \texttt{vann:preferredNamespacePrefix} is \texttt{tro}, which is registered at the prefix.cc resource\footurl{http://prefix.cc/tro}.

\section{Conclusion and future work}\label{sec:conclusion-future-work}

The Transparent Relations Ontology (TRO) provides a vocabulary for describing potential conflicts of interest. Given that data pertaining to such conflicts is normally found in isolated resources with heterogeneous formats, a vocabulary like TRO is needed to unify it, as exemplified by the use case of Section \ref{sec:usingTRO}. 

TRO is in its first steps, but the best practices followed in the ontology development should help in on-boarding new developers: Additionally, the future development of TRO will also include technical improvements like new ROBOT tests for a higher-quality artefact production.

Since TRO's design has been inspired by data from existing projects, it is expected that its use will grow. This growth will accommodate new use cases organically, as more activists are interested in this domain, specially in current times of societal interest for corruption cases. This interest will result, ultimately, in an improved government transparency.

\section{Acknowledgements}\label{sec:acknowledgements}

The author would like to thank the valuable comments from the participants in the Bilbao Data Lab\footurl{https://bilbaodatalab.wikitoki.org/} project, which is part of WikiToki\footurl{https://wikitoki.org/en/about/}.

\bibliography{main}

\bibliographystyle{abbrv}

\end{document}